\documentclass[10pt, aps, prl, superscriptaddress, %nobibnotes, nofootinbib, 
               amsmath, amssymb, twocolumn,
               preprintnumbers, showpacs,
               raggedbottom, floatfix]{revtex4-1}

\usepackage[arXiv]{my_paper}                     % Use this for the arXiv

\usepackage{my_acronyms}                         % acronyms for this paper.

\usepackage[utf8]{inputenc}

\usepackage{sparklines}

\begin{document}

\title{Strength of the Vortex-Pinning Interaction from Real-Time Dynamics}

\author{Aurel Bulgac}
\affiliation{Department of Physics, University of Washington, Seattle,
  Washington 98195--1560 USA}

\author{Michael McNeil Forbes}
\affiliation{Institute for Nuclear Theory, University of Washington,
  Seattle, Washington 98195--1550 \upsc{USA}}
\affiliation{Department of Physics, University of Washington, Seattle,
  Washington 98195--1560 \upsc{USA}}

\author{Rishi Sharma} \affiliation{\upsc{TRIUMF}, Vancouver, British Columbia,
  \upsc{V6T 2A3}, Canada}

\date{\today}

\begin{abstract}\noindent
  We present an efficient and general method to compute vortex-pinning
  interactions -- which arise in neutron stars, superconductors, and trapped
  cold atoms -- at arbitrary separations using real-time dynamics. This method
  overcomes uncertainties associated with matter redistribution by the vortex
  position and the related choice of ensemble that plague the typical approach
  of comparing energy differences between stationary pinned and unpinned
  configurations: uncertainties that prevent agreement in the literature on the
  sign and magnitude of the vortex-nucleus interaction in the crust of neutron
  stars.  We demonstrate and validate the method with Gross-Pitaevskii--like
  equations for the unitary Fermi gas, and demonstrate how the technique of
  adiabatic state preparation with time-dependent simulation can be used to
  calculate vortex-pinning interactions in fermionic systems such as the
  vortex-nucleus interaction in the crust of neutron stars.
\end{abstract}
\preprint{%
  \mysc{nt\ampersat uw-13-09},
  \mysc{int-pub-13-005}}

\maketitle
\glsresetall
\glsunset{BCS}

\lettrine{V}{ortex-pinning interactions} play an important role in the dynamics
of various condensed superfluid systems such as
superconductors~\cite{Shapoval:2010}, trapped cold-atom gases~\cite{Tung:2006},
and possibly neutron stars~\cite{Link:1999}, where the angular momentum carried
by vortices can have an observable impact.  For example, pulsar glitches --
sudden increases in the rotation frequencies of neutron stars -- are
theorized~\cite{Anderson:1975} to arise from a sudden macroscopic unpinning of
vortices. In equilibrium, the superfluid and nonsuperfluid components of a
pulsar rotate at the same angular frequency. The pulsar loses angular momentum
through magnetic radiation, and the crust slows down gradually, reducing the
pulsation rate. To maintain equilibrium, the superfluid must also release
angular momentum by diluting the vortex concentration, but the presence of
pinning sites (nuclei, lattices sites, defects, etc.) may arrest the vortex
motion; stress would build until a large number of vortices rapidly unpin,
dilute, and transfer their angular momentum to the crust, rapidly increasing
in the pulsation rate -- the glitch.

Despite almost 40 years, the feasibility of this mechanism is still poorly
understood.  The conventional picture has the angular momentum stored by the
neutron superfluid in the crust, with pinning provided by nuclei held in a
lattice by the electrostatic (Coulomb) interaction. (Dilute neutron matter is
well approximated~\cite{Carlson:2012} by the same \gls{UFG} produced in
cold-atom experiments~\cite{Zwerger:2011}.)  Pinning may also occur on flux
tubes~\cite{Ruderman:1998} or due to vortex tangles~\cite{Peralta:2006}. Recent
results suggest that the crustal neutrons may not support enough angular
momentum to explain observed pulsar glitches~\cite{Andersson:2012,
  *Chamel:2013a}, in which case the interaction between neutron superfluid
vortices and proton flux tubes in the outer core~\cite{Link:2012qe} or quark
matter phases in the core may play a
role~\cite{Bowers:2002xr,*Mannarelli:2007bs}. In either case, a reliable
technique for calculating vortex-pinning interactions is key.  Here we present a
dynamical method for determining the sign and strength of vortex-pinning forces
in superfluids, and demonstrate that this method can be directly applied to
unambiguously calculate the vortex-nucleus interaction using \gls{TDDFT} for
nuclear matter.

% Full-width figures use figure* and need to come on the prior page in the .tex
% file.
\begin{figure*}[t]
  \includegraphics[width=\textwidth]{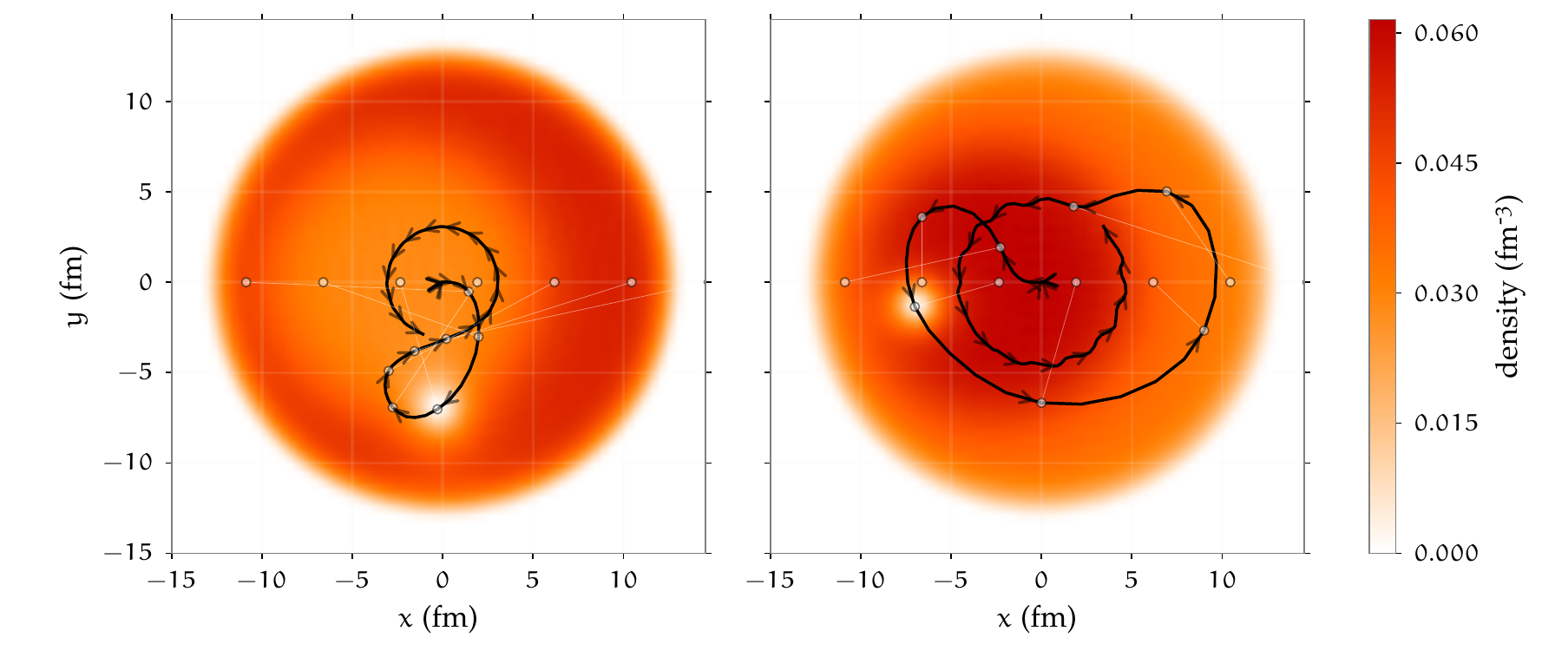}
  \caption{\label{fig:deflection}(colour online) %
    Deflection of a vortex in the \acrshort{ETF} model of trapped dilute neutron
    matter as a \acrshort{UFG} by a repulsive (left panel) and attractive (right
    panel) pinning potential $V_{\text{pin}}(r) = \pm
    \SI{3.5}{MeV}/[1+\exp(r/\si{fm}-7.5)]$ moving on a straight line from left
    to right at a constant subsonic velocity $v \approx 0.1c_s$. The trajectory
    of the vortex is shown by the (black) curve and the relative separation
    vector between the pinning site, and the vortex core is shown as thin
    (white) lines for select times connecting the corresponding dots on the
    trajectories.  Initially the potential displaces the bulk superfluid,
    carrying the vortex to the right/left.  Once the potential overlaps with the
    vortex, the vortex rapidly moves down/up -- (almost) perpendicular to the
    force. In the frame shown on the left, the pinning site is just to the left
    of the centre ($x\approx \SI{-2.5}{fm}$) and the vortex is moving (almost)
    perpendicular along the edge of the pinning potential.  After the potential
    has passed through, the vortex orbits in a counterclockwise circle
    direction due to boundary effects from the trap that can be quantitatively
    described in this sharp, flat trap by placing an image vortex outside of the
    potential to cancel the tangential current at the boundary: this induces a
    counterclockwise superflow $v_s$ in Eq. ~\eqref{eq:HVI}. The geometry of
    the right simulation is such that the potential carries the vortex around
    almost the entire trap: This extended interaction allows the pinning
    potential to excite phonons in the system visible as ripples in the circular
    trajectory.}
\end{figure*}

Because of the importance of pinning on glitch phenomenology, several attempts
have been made to calculate the pinning force in nuclear matter from underlying
microscopic models. The earliest calculations used the condensation energy to
estimate the pining force~\cite{Anderson:1975, Alpar:1977, *Alpar:1984}.  In
Refs.~\cite{Epstein:1988, *Link:1991}, the \gls{GL} framework was used to give a
detailed picture of the (un)pinning process: their calculation includes an
estimate of the energy as a function of displacement allowing for an estimate of
the pinning force.  The next advance was the use of a local density
approximation~\cite{Pizzochero:1997} with Wigner-Seitz cells and
Gogny~\cite{Broglia:1994} and Argonne~\cite{Baldo:1990} interactions, which
gives a similar density-dependant pattern of (un)pinning as~\cite{Link:1991}
but smaller by almost an order of magnitude.

Unlike vortices in weakly coupled \gls{BCS} superfluids, vortices in dilute
neutron matter (and the \gls{UFG}) displace a substantial amount of matter from
their core~\cite{Yu:2003a}.  Therefore, comparing ``energies'' of stationary
configurations with a nucleus on the core of a vortex and a nucleus away from
the vortex is confounded by a choice of ensemble: should one fix the number of
neutrons or the chemical potential in a finite simulation volume?

Computing stationary configurations is also computationally expensive --
especially given the high degree of precision required to render meaningful
energy differences -- and simulations to date have required a high degree of
symmetry. For example, recent self-consistent calculations~\cite{Donati:2004,
  Avogadro:2007, Avogadro:2008} using~\gls{HFB} functionals extract the pinning
energy of a vortex on a single nucleus using a cylindrical geometry. In
particular, the conclusion of~\cite{Avogadro:2007} that the pinning force is
repulsive (glitches would thereby require interstitial pinning) was questioned
by~\cite{Pizzochero:2007} but addressed in~\cite{Avogadro:2008}, while a
different set of calculations using the local density approximation suggests
that pinning is attractive over a substantial region in the inner
crust~\cite{Donati:2004, Donati:2003, *Donati:2006}.  Moreover, nearby vortices
and the Casimir effect can significantly polarize a nucleus -- an effect absent
in simple cylindrical geometries -- dramatically changing the nature of the
nuclear pinning sites and disrupting the regularity of the nuclear
lattice~\cite{Bulgac:2001, *Bulgac:2001E, *Magierski:2002a, *Magierski:2003}.

Characterizing the nuclear-pinning interaction will thus require fully 3\mysc{D}
(unconstrained by symmetries) self-consistent calculations using realistic
nuclear functionals.  Highly accurate asymmetric stationary states in full
3\mysc{D} are currently not feasible (these require a full diagonalization of
the single-particle Hamiltonian), but \gls{TDDFT} algorithms can be applied to
the unconstrained 3\mysc{D} problem (which requires only applying the
Hamiltonian), and scale well to massively parallel supercomputers for both cold
atoms and nuclei, as has been demonstrated in~\cite{Bulgac:2011b, *Stetcu:2011,
  *Bulgac:2011c}.  We now present a qualitatively new approach for calculating
vortex-pinning interactions, unencumbered by the aforementioned issues,
utilizing only real-time dynamics.

\begin{figure}[tp]
  \includegraphics[width=\columnwidth]{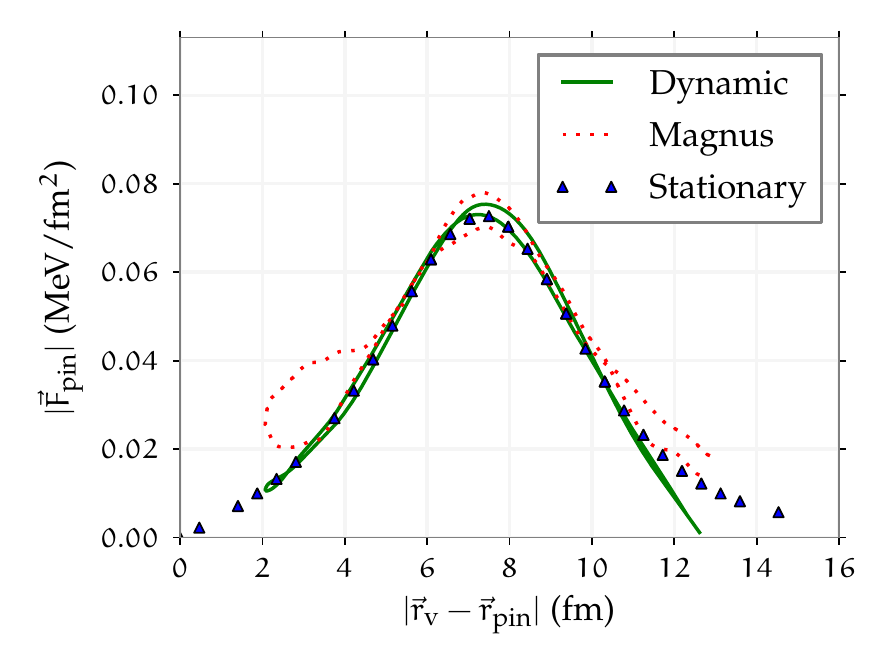}
  \caption{\label{fig:force} (color online) Here we demonstrate consistency in
    dynamically extracting a vortex-pinning force.  We use the nuclear pairing
    potential~\cite{Broglia:1994} $V_{\text{pin}}(r) =
    \SI{0.75}{MeV}/[1+\exp(r/\si{fm}-7.5)]$ at densities $\rho \sim
    \SI{0.045}{fm^{-3}} \approx 0.28\rho_{\text{sat}}$.  The triangular (blue)
    points come from the computationally expensive ``stationary'' method, while
    the solid (green) curve comes from using the ``dynamic'' real-time evolution
    analogous to that shown on the left panel of Fig.~\ref{fig:deflection}.  The
    dotted (red) curve shows the Magnus estimate for the force~\eqref{eq:HVI}
    using a Thomas-Fermi approximation for $\rho_s$ and estimating $\vv_s$ from
    the image vortex~\cite{PS:2002}. The double curves come from the pinning
    site moving in then out.}
\end{figure}

The idea, similar to the Stern-Gerlach experiment, is to observe how a vortex
moves when approached by a nucleus.  To zeroth order, the sign of the
interaction is determined qualitatively by the direction of the motion
(Fig.~\ref{fig:deflection}); with a more careful inspection, one can extract the
force-separation relationship $F(r)$ (Fig.~\ref{fig:force}).

We validate our procedure using a dynamical \gls{ETF} model~\cite{Kim:2004,
  Salasnich:2008b, *Salasnich:2008, *Salasnich:2008E, *Salasnich:2011,
  *Salasnich:2012, Forbes:2012, Forbes:2012b}, equivalent to a \gls{GPE} for
bosonic ``dimers'' $m_B = 2m$ of fermionic pairs, with an equation of state
$\mathcal{E}(n) \propto \xi \rho^{5/3}$ characterized by the Bertsch parameter
$\xi\approx 0.37$ tuned to consistently fit both \gls{QMC} and experimental
results~\cite{Forbes:2012}. Despite the computational simplicity of the
\gls{ETF} model, it has been demonstrated to quantitatively reproduce a range of
low-energy dynamics of both \gls{UFG} experiments~\cite{Salasnich:2012} and
fermionic \gls{DFT} simulations~\cite{Forbes:2012b}. The \gls{UFG} should also
qualitatively model the dilute neutron superfluid in the crust of neutron
stars~\cite{Carlson:2012} due to the large neutron-neutron scattering length
$a_{nn}\approx\SI{-18.9}{fm}$~\cite{Chen:2008}.  Thus, by using a physically
motivated model of the nuclear pairing potential~\cite{Broglia:1994}, we
anticipate that these \gls{ETF} calculations will provide a fairly good
approximation of future fermionic \gls{TDDFT} simulations.

To gain some intuition for the vortex-nucleus interaction, consider the
phenomenological \gls{HVI} equation (see~\cite{Thompson:2011} for a discussion)
for a vortex in 2\mysc{d}:
\begin{gather}
  \label{eq:HVI}
  M\ddot{\vr}_v - \vf_{qp} = \rho_s\vkappa\times(\dot{\vr}_v - \vv_s) + \vF_v.
\end{gather}
Here, $\vr_v$ is the position of the vortex, the force $\vF_v$ is per unit
length along the vortex, $\rho_s$ is the number density of the ``background''
superfluid, $\vkappa = 2\pi\hbar\uvect{z}$ is the quantized vortex circulation,
and $\vv_s$ is the ``background'' superfluid velocity.  This equation should
only be taken as an intuitive guide since terms on the left-hand side are
ill-defined. The ``mass of the vortex'' $M$, for example, depends strongly on
the way it is measured~\cite{Thouless:2007}, and the force $\vf_{qp}$ due to
excited phonons has significant memory effects.

For slowly accelerating vortices, the contribution from the term proportional to
$\ddot{\vec{r}}_v$ is small. Furthermore, if the vortex and pinning site move
sufficiently slowly, phonons are not excited ($\vec{f}_{qp} = 0$), and we can
ignore the entire left-hand side of Eq.~\eqref{eq:HVI}~\cite{Vinen:2002}.  This
leaves the well-established Magnus relationship $\rho_s\vkappa\times(\dot{\vr}_v
- \vv_s) \approx -\vF_v$ relating the force $\vF_v$ applied to the vortex and
its perpendicular velocity $\dot{\vr}_v$ relative to the background superfluid
velocity $\vv_s$. Thus, by observing the dynamical deflection of a vortex from a
nuclear pinning site, one can directly extract the direction and approximate
magnitude of the vortex-nucleus force without requiring a subtle subtraction of
energies.

In small systems, the Magnus relation can only be used to estimate the magnitude
of the force since the superfluid density $\rho_s$ and velocity $v_s$ are not
precisely defined, though reasonable estimates can be obtained.  With an
external pinning potential $V_{\text{pin}}(\vr_{\text{pin}}-\vx)$, however, one
can directly and unambiguously calculate the force on the pinning site:
\begin{gather}
  \label{eq:force}
  \vF_{\text{pin}} = - \int \d^3{x}\; 
  \pdiff{V_{\text{pin}}(\vr_{\text{pin}}-\vx)}
        {\vr_{\text{pin}}} \rho(\vx).
\end{gather}
In the nuclear context where neutrons are present in the both the pinning site
(the nucleus) and the superfluid medium, the force can be obtained in two ways:
1) Eq.~\eqref{eq:force} can be directly applied to a Coulomb potential
($V_{\text{pin}}$) that couples to the proton charge density ($\rho$) -- this
will be the force that the vortex exerts on the nuclear lattice -- or, 2) one
can estimate the force using Newton's law
$\vF_{\text{pin}}=m_{\text{pin}}\vect{a}_{\text{pin}}$ for a dynamic pinning
site comprising protons and entrained neutrons.  The position of the pinning
site can be unambiguously defined as the \gls{CM} of the protons, and the
effective mass $m_{\text{pin}}$ can be estimated~\cite{Magierski:2004,
  *Magierski:2004a}.

What remains is to prepare the initial conditions with a vortex and nucleus
interacting at various distances.  The traditional self-consistent approach
requires diagonalizing $N\times N$ matrices ($N = N_xN_yN_z$) which takes
$\order(N^3)$ operations.  This is not feasible for realistic $N\sim 10^6$, as
\emph{each iteration} would required a day of supercomputing wall time.
Instead, one can use adiabatic state preparation~\cite{Pfitzner:1994,
  Bulgac:2013c} which takes $\order(N^2\log N)$ operations. The idea is to
adiabatically evolve in real time a state of some solvable system to a desired
initial state in the system of interest. For example, starting with a
noninteracting (Bose) gas trapped in a harmonic potential $V_{\mysc{ho}}(r) =
m_B\omega^2r^2/2$, we can form either the ground state $\Psi_{\mysc{gs}} \propto
\exp(-m_B \omega r^2 /2 )$, or an exact vortex ``Landau level'' $\Psi_\delta
\propto (x + \I y - \delta) \exp(-m_B \omega r^2 /2\hbar )$ (stationary in a
rotating frame) with angular momentum $l_z = N\hbar/(1 +
m_B\omega\delta^2/\hbar)$ where $\delta$ is the displacement of the vortex node
from the centre of the harmonic trap.  From this exact noninteracting state we
adiabatically evolve the system to an interacting state in the desired trapping
potential $V_{\text{trap}}$ by simultaneously switching on the interaction
$s\xi$ and interpolating the trapping potentials $V_t = (1-s)V_{\mysc{ho}} +
sV_{\text{trap}}$ where $s=s(t/T)$ is a smooth $C^{\infty}$ switching function
that goes from $0$ to $1$ over a characteristic time $T$ chosen to be longer
than any intrinsic time scale in the system:
\begin{gather}
  s\left (\frac{t}{T}\right ) =
  \frac{1}{2} + \frac{1}{2}\tanh \left[ 
    \alpha\tan \left ( \frac{\pi t}{T} - \frac{\pi}{2} \right ) \right ]
  = \sparkSS
\end{gather}
%(The parameter $\alpha$ controls the smoothness and we find empirically that
%values $1 < \alpha\lessapprox 6$: \sparkSS\ work well.)
From $\Psi_{\mysc{gs}}$ we can generate the ground state, and from
$\Psi_{\delta=0}$ we can generate a single vortex in the centre of the trap,
both to high precision.  The adiabatic state preparation can be significantly
accelerated by introducing a ``quantum friction'' term to remove phonon
noise~\cite{Bulgac:2013c}.  With this combined approach, one can
efficiently produce almost any desired initial state with less than a day of
supercomputing wall time.

To accurately measure the vortex-pinning interaction, one can choose as a final
potential $V_T = V_\text{trap} + V_\text{pin}$: an axially symmetric trap of
suitably flat bottom and a pinning potential in the center.  By generating a
configuration with a vortex orbiting in a circle at radius $r$, we can use
Eq.~\eqref{eq:force} to calculate the force exerted on the centrally located
pinning potential: axial symmetry ensures that this is precisely the
vortex-pinning force at separation $r$.  We use this procedure within the
\gls{ETF} model to accurately calculate the ``stationary'' (in a rotating frame)
vortex-pinning interaction shown in Fig.~\ref{fig:force}.

The present demonstration has been limited to quasi-2\mysc{d} simulations.  The
procedure will work just as well in fully 3\mysc{d} simulations.  New effects
such as the bending of a vortex line when approached by a pinning site can just
as easily be analyzed: the vortex line will either be repelled by the pinning
site -- bowing out to avoid it -- or will be sucked in.    We
have considered here only moving the pinning site, but one could also consider
manipulating parts of the vortex with pinning potentials, dragging the pinned
vortex along a trajectory instead.  In simulations with realistic nuclei, the
vortex-nucleus interaction will also excite and deform the nucleus --
significantly affecting the vortex-nucleus interaction.  It is conceivable also
that a vortex lines could break and attach to various nuclear defects like rods
or plates: the dynamics of such broken vortex lines may also play a important
part in explaining neutron star glitches.

\newcommand{\cross}{\times}
We close with a brief analysis of the time evolution.  The complex scalar field
$\Psi$ obeys an evolution equation of the form
\begin{gather}
  \I \hbar \dot{\Psi} =    \left(
    -\frac{\hbar^2\vnabla^2}{2m_B} + V_{\text{eff}}[\Psi]
  \right)\Psi
\end{gather}
where $V_{\text{eff}}[\Psi]$ is an effective interaction that depends
nonlinearly on $\abs{\Psi}$ and on the external trapping and pinning
potentials.  Consider the quasi-two-dimensional problem where coordinates may
be expressed as complex numbers $z=x+\I y $: A singly-wound vortex at location
$z_v$ may be described by the field $\Psi(z) = (z - z_v)f(z)$, where $f(z)$ is a
smooth complex-valued function that we assume has no roots in the immediate
vicinity of the vortex.  The evolution equations may then be expressed as
follows
\begin{gather}
  \I\hbar\dot{f} + \frac{\hbar^2\vnabla^2f}{2m_B} - V_{\text{eff}}[\Psi]f
  =
  \frac{\I\hbar\dot{z}_v f - \hbar^2(\partial_xf + \I\partial_yf)/m_B}{z-z_v}.
\end{gather}
The left side is smooth; hence, the pole on the right side must
cancel with a root in the numerator, giving us an explicit expression for the
vortex velocity 
\begin{gather}
  \dot{z}_v = [\dot{\vr}_v]_x + \I[\dot{\vr}_v]_y  
     =\left . \hbar \frac{-\I\partial_xf + \partial_yf}{m_B f}\right |_{z=z_v}.
\end{gather} 
This expression can be written as an exact ``Magnus'' relation
\begin{gather}\label{eq:MagnusExact}
  \vkappa \times (\dot{\vr}_v - \vv) 
  = \left.\frac{\hbar^2\vnabla{\rho}}{2m_B \rho}\right|_{z=z_v}
  \intertext{where the local ``superfluid velocity'' $\vv$ is}
  \begin{aligned}
    f &= \sqrt{\rho}e^{\I\phi}, & \vv &=\vnabla\phi.
  \end{aligned}
\end{gather}
The meaning of $\rho_s$ in the \gls{HVI} equations is not clarified since $\rho$
cancels in Eq.~\eqref{eq:MagnusExact}. Unfortunately, although $\vv$ is
precisely defined and corresponds to $\vv_s$ in some situations, one cannot
generally make the correspondence $\vv_s \equiv \vv$.  In particular, doing so
yields results that differ by as much as 50\% in Fig.~\ref{fig:force}.

The quantities appearing in Eq.~\eqref{eq:HVI} are related to long-range
momentum transfers and boundary effects, and one must thus be content with
reasonable estimates for $\rho_s$ and $v_s$, for example, from the average
behaviours of the relevant quantities near but outside of the vortex core. As
Fig.~\ref{fig:force} demonstrates, however, the Magnus relation is suitable for
extracting the sign and magnitude of the interaction.  This also provides an
explicit check that the force evaluated using our procedure is what appears on
the right side in Eq.~\eqref{eq:HVI} governing the vortex dynamics.

\glsresetall
\paragraph{Conclusion:}
We have described how to use \gls{TDDFT} to efficiently and unambiguously
calculate vortex-pinning interactions from real-time dynamical simulations of
superfluid systems.  We have demonstrated with an \gls{ETF} model of the
\gls{UFG} that this approach can be applied to calculate the vortex-nucleus
interaction using nuclear \glspl{TDDFT} to model the crust of neutron stars.
While we considered only quasi-2\mysc{d} systems here, the size of the problem,
the magnitude and accuracy of the force extraction, and the use of pure
real-time dynamics ensure that full 3\mysc{d} simulations of realistic fermionic
\glspl{TDDFT} are possible. With available resources~\cite{Bulgac:2011b,
  *Stetcu:2011, *Bulgac:2011c} one can simulate both finite and infinite nuclear
systems in simulation boxes of the order of $80^{3}\,$\si{fm^{-3}} for up to
$10^{-19}\,$\si{s}.  A resolution to the puzzle of pulsar glitches will require
more than just extracting the vortex-nucleus interaction, but with this
real-time method, this crucial step will soon be within reach.

These techniques can also be directly applied to systems of trapped ultracold
atoms in a variety of geometries, for example, to explore vortex pinning on
optical lattices. In particular, the close approximation of the neutron
superfluid by the \gls{UFG} suggests that cold-atom experiments might also be
able to shed light on the glitching puzzle.

\providecommand{\MMFGRANT}{\mysc{de-fg02-00er41132}}
\begin{acknowledgments}
  We thank S.~Reddy and D.~Thouless for useful discussions.  This material is
  based upon work supported by the U.~S.~Department of Energy under Award
  Numbers \mysc{de-fg02-97er41014} and \MMFGRANT.
\end{acknowledgments}

%\bibliographystyle{apsrev4-1}
%\bibliography{local,master}


\end{document}